\documentclass[preprint,showpacs,amssymb,aps,prd]{revtex4}

\begin{document}

\def\diagram#1{{\normallineskip=8pt
       \normalbaselineskip=0pt \matrix{#1}}}

\def\diagramrightarrow#1#2{\smash{\mathop{\hbox to
.8in{\rightarrowfill}}
        \limits^{\scriptstyle #1}_{\scriptstyle #2}}}

\def\diagramleftarrow#1#2{\smash{\mathop{\hbox to .8in{\leftarrowfill}}
        \limits^{\scriptstyle #1}_{\scriptstyle #2}}}

\def\diagramdownarrow#1#2{\llap{$\scriptstyle #1$}\left\downarrow
    \vcenter to .6in{}\right.\rlap{$\scriptstyle #2$}}

\def\diagramuparrow#1#2{\llap{$\scriptstyle #1$}\left\uparrow
    \vcenter to .6in{}\right.\rlap{$\scriptstyle #2$}}

%
\title{Anomaly in conformal quantum mechanics:
\\
From molecular physics to black holes}

\author{ Horacio E. Camblong}

\affiliation{
Department of Physics, University of San Francisco, San
Francisco, California 94117-1080, USA}

\author{Carlos R. Ord\'{o}\~{n}ez}

\affiliation{
Department of Physics, University of Houston, Houston,
Texas 77204-5506, USA
\\
World Laboratory Center for Pan-American Collaboration in Science and
Technology,
\\
University of Houston Center, Houston, Texas 77204-5506, USA
}

\begin{abstract}
A number of physical systems exhibit a particular form of asymptotic 
conformal invariance: within a particular domain of distances, 
they are characterized by a long-range conformal interaction
(inverse square potential), the apparent absence of dimensional scales, 
and an SO(2,1) symmetry algebra. 
Examples from molecular physics to black holes are provided and discussed
within a unified treatment.
When such systems are physically realized in the 
appropriate strong-coupling regime, 
the occurrence of quantum symmetry breaking is possible. This anomaly is 
revealed by the 
failure of the symmetry generators to close the algebra in a manner shown to 
be independent of the renormalization procedure.
\end{abstract}
\maketitle

\section{Introduction}
\label{sec:introduction}

An anomaly is defined as the symmetry breaking of a classical 
invariance at the quantum level.
This intriguing phenomenon has played a crucial role in theoretical physics since its 
discovery in the 1960s~\cite{tre:85}.
In addition to its use in particle phenomenology of the 
standard model~\cite{don:92} and its extensions,
it has been a fruitful tool for the study of conformal invariance in
 string theory~\cite{schwartz_witten}.

Surprisingly,
the presence of
 an infinite number of degrees of freedom
does not appear to be a prerequisite for the emergence of anomalies.
This fact was first recognized 
within a model with conformal invariance:
the two-dimensional contact interaction
 in quantum mechanics~\cite{jackiw:91-beg}.
In conformal quantum mechanics,
a physical system is classically
invariant under the most general 
combination of the following time reparametrizations:
time translations,
generated by the Hamiltonian $H$;
 scale transformations, generated by the
 dilation operator
$
D
\equiv
tH
- 
\left( {\bf p} \cdot {\bf r}
+  {\bf r} \cdot {\bf p}
\right)/4$; and 
translations of reciprocal time,
generated by the 
special conformal operator
$
 K
\equiv
2t D -
t^{2}H
+ 
m r^{2}/2
$.
These generators define
the noncompact
SO(2,1) $\approx$ SL(2,R) Lie algebra~\cite{wyb:74} 
\begin{equation}
[D,H]
= - i \hbar H
 \;  ,
\; \;  
[K,H]
= - 2 i \hbar D
\;  ,
\; \;  
[D, K]
=  i \hbar K
\;  .
\label{eq:naive_commutators}
\end{equation}
This symmetry algebra has also been recognized in
 the free nonrelativistic particle~\cite{niederer-hagen:nr-SO(21)},
 the inverse square 
potential~\cite{jackiw:72,alfaro_fubini_forlan:76},
the magnetic
monopole~\cite{jackiw:80},  the magnetic vortex~\cite{jackiw:90},
and  various nonrelativistic quantum field 
theories~\cite{niederer-hagen:nr-SO(21),jackiw:nr-SO(21),bergman:nr-SO(21)}.
Furthermore,
conformal quantum mechanics has 
been fertile ground for the study of 
singular potentials and
renormalization,
using Hamiltonian~\cite{gup:93, camblong:isp-dt,beane:00}
as well as path integral methods~\cite{pi_collective}.
Most importantly,
a variety of physical realizations of conformal quantum mechanics
have been recently identified,
as discussed in the next section. 

The main goals of this paper are 
(i)  to illustrate the relevance 
of conformal quantum mechanics
for several physical problems,
from  molecular physics to black holes, and
(ii)
to show the details  of the breakdown of the 
commutator algebra~(\ref{eq:naive_commutators})
for the  
long-range conformal interaction.
In Sec.~\ref{sec:physical_applications} 
we introduce a number of examples that can be regarded as 
physical realizations of conformal quantum mechanics.
In Sec.~\ref{sec:CA_UV_physics}
we show that the origin of the anomaly can be traced 
to the short-distance singular behavior of the conformal interaction.
In Sec.~\ref{sec:regularization_renormalization}
we introduce a generic class of real-space regulators,
within the philosophy of the effective-field theory program. 
In Sec.~\ref{sec:CA_computation} we compute the anomaly for the 
regularized theory and show that it is independent of the 
details of the ultraviolet 
physics, and in Sec.~\ref{sec:renormalization_frameworks}
we comment on various renormalization frameworks.
After the conclusions in Sec.~\ref{sec:conclusions},
we summarize a number of technical results:
a derivation of the anisotropic generalization
of the conformal long-range interaction
(Appendix~\ref{sec:anisotropic_ISP_and_dipole_bound_anions});
a study of interdimensional dependence
(Appendix~\ref{sec:dimensionalities_and_interdimensional});
a derivation of the near-horizon properties of black holes 
(Appendix~\ref{sec:NH_conformal_behavior});
and a derivation of useful
integral identities (Appendix~\ref{sec:generalized_Lommel_integrals}).

\section{Relevant Physical Applications}
\label{sec:physical_applications}

In recent years,  
diverse examples of systems have been studied from the viewpoint
of the conformal algebra~(\ref{eq:naive_commutators}),
assumed to be a representation of
an approximate symmetry
within specific scale domains.
In the applicable conformally invariant domain, the relevant physics is
described by a $d$-dimensional  {\em effective\/} Hamiltonian
\begin{equation}
H= \frac{ p^{2}}{2m}  - \frac{g}{r^{2}}
\; ,
\label{eq:ISP_Hamiltonian_unregularized}
\end{equation}
which involves
a long-range conformal interaction; or, alternatively,
by its anisotropic counterpart
\begin{equation}
H= \frac{ p^{2}}{2m}  - \frac{g}{r^{2}} \, F({\bf \Omega})
\; ,
\label{eq:ISP_Hamiltonian_unregularized_anisotropic}
\end{equation}
where  ${\bf \Omega}$ stands for the angular variables and 
$F({\bf \Omega})$ is a generic anisotropy
factor that accounts for the angular dependence.
Equation~(\ref{eq:ISP_Hamiltonian_unregularized_anisotropic}) 
is discussed in Appendix~\ref{sec:anisotropic_ISP_and_dipole_bound_anions}.

In the problems considered below,
$\lambda = 2m g/\hbar^{2}$
is the dimensionless form of the coupling constant and
$\nu = (d-2)/2$; furthermore, the choice $\hbar = 1 =  2 m$ 
will be made for the problem involving black holes.
In all cases,
the strong-coupling regime~\cite{camblong:isp-dt}
is defined by the condition $g \geq g^{(*)}$,
with a critical dimensionless coupling 
$\lambda^{(*)}
\equiv \lambda_{l}^{(*)} = (l + \nu)^{2}$ (for 
angular momentum $l$), when the Hamiltonian 
model~(\ref{eq:ISP_Hamiltonian_unregularized})
is adopted~\cite{camblong:isp-dt}. In addition,
in the strong-coupling regime, as deduced in 
Sec.~\ref{sec:CA_UV_physics},
an uncontrolled oscillatory behavior of
the Bessel functions of imaginary order $i \Theta$
makes the conformal system singular
and regularization is called for.
The characteristic parameter
 $ \Theta = \sqrt{\lambda - (l+\nu)^{2}}$
strictly corresponds to the Hamiltonian~(\ref{eq:ISP_Hamiltonian_unregularized});
in physical applications,
such as those of Secs.~\ref{sec:dipole_bound_anions}, \ref{sec:NH_black_holes},
and \ref{sec:Efimov_effect}, we will define
\begin{equation}
\Theta  \equiv \Theta_{\rm eff}
= \sqrt{\lambda_{\rm eff} - 
\lambda_{\rm eff}^{(*)} }
\; ,
\label{eq:Theta_ISP_parameter}
\end{equation}
which will turn out to be crucial in parametrizing the anomalous physics 
of the conformal system in the presence of symmetry breaking.
In discussing these realizations,
we will explicitly use a subscript to
emphasize the effective nature 
of the parameter of Eq.~(\ref{eq:Theta_ISP_parameter})---as arising
from a reduction framework.
The same notational convention will apply to the dimensionality ($d_{\rm eff}$).
As shown in Appendix~\ref{sec:dimensionalities_and_interdimensional},
even when interdimensional equivalences are introduced,
the value of the parameter~(\ref{eq:Theta_ISP_parameter})
is a {\em dimensional invariant\/}.

\subsection{Dipole-bound anions and anisotropic conformal interaction}
\label{sec:dipole_bound_anions}

The three-dimensional ($d_{\rm eff}=3$ or $\nu_{\rm eff} =1/2$)  
interaction between an electron  (charge $Q=-e$) 
and a polar molecule (dipole moment ${\bf p}$)
was the first physical application 
to be recognized as a realization of this 
anomaly~\cite{molecular_dipole_anomaly}. 
When the molecule is modeled as a  point dipole, this interaction 
can be effectively described 
with an anisotropic long-range conformal interaction of the 
form~(\ref{eq:ISP_Hamiltonian_unregularized_anisotropic}):
$V({\bf r}) = -g \cos \theta /r^{2}$,
in which  the polar angle $\theta$ is subtended
 from the direction of the dipole moment. 
For this potential, the dimensionless coupling is
$
\lambda = - 2 m \, K_{e} p Q/\hbar^{2}=
p/p_{0}
$,
with $m$ being the reduced mass of the system and $K_{e}$ the
electrostatic constant. 
Thus, the relevant scale  for phenomenological analyses is provided by
  $p_{0} \approx 1.271 $
$D$ (where $D $ stands for the
debye).

As shown in Appendix~\ref{sec:anisotropic_ISP_and_dipole_bound_anions},
in some sense, the generic anisotropic conformal 
interaction~(\ref{eq:ISP_Hamiltonian_unregularized_anisotropic})---of 
which the electron-molecule interaction is a particular 
case---can be reduced to 
an {\em effective isotropic\/} conformal interaction for the zero 
angular-momentum channel
[see Eq.~(\ref{eq:effective_reduction_anisotropic-->>isotropic})];
this corresponds to an effective Hamiltonian
of the type~(\ref{eq:ISP_Hamiltonian_unregularized}),
with an appropriate effective coupling $\lambda_{\rm eff}$.
More precisely, this equivalence is achieved, 
after  separation of variables in spherical coordinates,
at the level of the radial equation.
In addition,
the corresponding 
value of  $\lambda_{\rm eff}$ is identical to the
eigenvalue $\gamma$ of the angular equation,
which  is a function of the dipole coupling $\lambda$.
The effective conformal parameter~(\ref{eq:Theta_ISP_parameter})
becomes
\begin{equation}
\Theta_{\rm eff} 
=
\sqrt{ \gamma   - \nu_{\rm eff} ^{2}}
\;  ,
\label{eq:Theta_effective_dipole}
\end{equation}
where $\lambda_{\rm eff}^{(*)}= \nu_{\rm eff}^{2}$
for each eigenvalue $\gamma  $ of the angular equation.
When this outline is implemented,
according to the procedure of Ref.~\cite{molecular_dipole_anomaly} 
or its generalization of
Appendix~\ref{sec:anisotropic_ISP_and_dipole_bound_anions},
the existence of a critical dipole moment  $p^{(*)} $
for binding is predicted; the order of magnitude of its
``conformal value,''
 $\lambda^{(*)} \approx 1.279$,
or $p^{(*)} \approx 1.625$ $ D$,
has been verified in numerous experiments~\cite{mea:84,des:94}.
In particular, when binding occurs,
extended states known as dipole-bound anions are 
formed. These conclusions have also been 
confirmed by detailed {\em ab initio\/}
calculations~\cite{mea:84,des:94} and by 
studies that incorporate the effects of rotational degrees
of freedom~\cite{dipole_rotational}, which
also modify slightly  the value of $p^{(*)} $.

In short,
the central issue in this analysis---also shared by the other physical
realizations discussed in this paper---is
the existence of a conformally invariant
domain whose ultraviolet boundary leads to the anomalous
emergence of bound states via renormalization. 
As a result, these states
break the original conformal symmetry of the model
and modify the commutators~(\ref{eq:naive_commutators}),
as we will show in the next few sections.
This 
simple fact alone captures the essence 
of the observed critical
dipole moment in polar molecules
and leads to an analytical prediction 
for the energies of the conformal states,
as discussed in Sec.~\ref{sec:renormalization_frameworks}
and Appendix~\ref{sec:anisotropic_ISP_and_dipole_bound_anions}.

\subsection{Near-horizon black hole physics}
\label{sec:NH_black_holes}

 A generic class of applications of conformal quantum mechanics
arises from the near-horizon conformal 
invariance of black holes,  its impact on
their  thermodynamics~\cite{carlip_solodukhin:near_horizon},
and its extension to superconformal 
quantum mechanics~\cite{superconformal}.
In particular, analyses  
based on the Hamiltonian~(\ref{eq:ISP_Hamiltonian_unregularized})
have been used to  explore horizon states~\cite{gov:BH_states,gupta:BH}
and to shed light on black hole thermodynamics~\cite{gupta:BH}.
Another class of current
applications~\cite{gupta:calogero} involves a many-body generalization
of Eq.~(\ref{eq:ISP_Hamiltonian_unregularized}):
the Calogero model,
which has also been directly linked to
black holes~\cite{calogero_black_holes}. 
These remarkable connections seem to confirm the conjecture that it is the 
horizon itself that encodes the quantum properties of a 
black hole~\cite{thooft:85}.

In this context,  we consider the spherically symmetric 
Reissner-Nordstr\"{o}m geometry 
 in $D$ spacetime dimensions,
whose metric
\begin{equation}
 ds^{2}
=
- f (r) \,  dt^{2}
+
\left[ f(r) \right]^{-1} \, dr^{2}
+ r^{2} \,
 d \Omega_{D-2}
\; 
\label{eq:RN_metric}
\end{equation}
is minimally
coupled to a
scalar field $\Phi(x)$ 
with action ($c=1$ and $\hbar=1$) 
\begin{equation}
S
=
-
\frac{1}{2}
\int
d^{D} x
\,
\sqrt{-g}
\,
\left[
g^{\mu \nu}
\,
\partial_{\mu} \Phi
\, 
\partial_{\nu} \Phi
+ 
m^{2} \Phi^{2}
\right]
\; .
\label{eq:massless_scalar_action}
\end{equation}
In Eq.~(\ref{eq:RN_metric}),
$ d \Omega_{D-2}$
stands for the metric on the unit $(D-2)$-sphere,
$
f (r)  =
1
-
2
\left( a_{M}/r \right)^{D-3}
 +
\left( b_{Q}/r \right)^{2(D-3)}
$,
and the lengths $a_{M}$ and $b_{Q}$ are determined from the
 mass $M$ and charge $Q$
of the black hole respectively~\cite{mye:86}.
In this approach, the conformal structure is
 revealed by a two-step procedure 
discussed in Appendix~\ref{sec:NH_conformal_behavior}
and consisting of:
(a) a reduction to an effective 
Schr\"{o}dinger-like equation,
to be analyzed in its frequency ($\omega$)
components;  (b)
the introduction of a near-horizon 
expansion in the variable
$x= r -r_{+}$ 
[with $r=r_{\pm}$ being the roots of $f(r)=0$, and $r_{+} \geq r_{-}$].
Two distinct scenarios emerge from this reduction:
the extremal and nonextremal cases,
when $r_{+} = r_{-}$
and $r_{+} \neq r_{-}$, respectively.
We will omit any discussion of the extremal case,
which is known to pose a number of conceptual difficulties
and is otherwise beyond the scope of the framework presented in this paper.
As for the 
 nonextremal case,
the following
facts arise from this reduction:

(i)
The ensuing effective problem is described by an interaction 
\begin{equation}
V(x) 
\stackrel{({\rm near \; horizon})}{\propto} 
-
x^{-2}
\; ,
\label{eq:ISP_BH_near_horizon}
\end{equation}
 which is conformally invariant
with respect to the near-horizon coordinate $x$.

(ii)  The effective Hamiltonian,
still being a $d$-dimensional
realization of the conformal interaction, does not have the usual
form corresponding to the radial part of a multidimensional 
Schr\"{o}dinger problem.
In particular, the angular momentum variables appear at a higher
order in the near-horizon expansion.

(iii)
 The coupling constant 
$\lambda_{\rm eff}$ is supercritical  for all nonzero frequencies.
This can be seen from
Eq.~(\ref{eq:conformal_interaction}),
which implies that
\begin{equation}
\Theta_{\rm eff} 
=
\frac{\omega}{  |f'(r_{+})| } 
\; .
\label{eq:coupling_ISP_BH_near_horizon}
\end{equation}
The conclusion from this procedure is that
the relevant physics occurs 
in the strong-coupling regime,
in  which the framework  discussed in this paper
can be applied.

\subsection{Other applications}
\label{sec:Efimov_effect}

While Secs.~\ref{sec:dipole_bound_anions}
and \ref{sec:NH_black_holes}
conform to the title of this paper,
applications in other areas of physics are also likely.
Among these, 
the Efimov effect~\cite{efimov_effect,jensen_Phys_Rep_Efimov} stands out.
This effect is expected to arise
in a three-body system with short-range interactions,
in which at least two of the two-body
subsystems have virtual or bound $s$-states 
near zero energy.
As in the case of the dipole-bound anions of 
Sec.~\ref{sec:dipole_bound_anions},
these are spatially extended and weakly bound states.
Unfortunately, the combination of phenomenological parameters 
needed to form these states, together with their weakly bound nature,
has defied experimental detection to date.
Nonetheless,
this effect is regarded as relevant
in the description of the 
three-body nucleon 
interaction~\cite{3_body_nucleon}.
The most outstanding feature of
these three-body interactions
in three spatial dimensions
 is the fact that these problems are reduced to
 an effective equation with a long-range
conformal interaction
in the  strong-coupling regime.
In terms of possible experimental detection, 
this effect is currently being studied for
 the description of various
 systems, including helium trimers 
and nuclear three-body halos~\cite{jensen_Phys_Rep_Efimov}.

The conformal nature of the effective interaction,
for the three-body systems
described above,
can be deduced as follows.
Typically,
one starts by introducing
hyperspherical coordinates with
hyperradius  $\rho \equiv r$, 
in a $d_{\rm eff}$-dimensional 
configuration space for the internal degrees of freedom;
 if the one-particle dynamics occurs in
a $d$-dimensional space,
then $ d_{\rm eff}
=2d$ for the internal dynamics of the 
three-body system
(as the total number of coordinates is $3d$, 
but $d$ of them are eliminated in favor 
of the center-of-mass coordinates). 
Consequently,
when a hyperspherical adiabatic expansion~\cite{hyperspherical_adiabatic}
is combined with 
a Faddeev decomposition of the wave function~\cite{faddeev_eqs},
a reduction to a  $d_{\rm eff}=2d$ realization of our conformal 
model~(\ref{eq:ISP_Hamiltonian_unregularized})
is obtained.
These conclusions can be gleaned from
the conformal nature of the effective adiabatic potentials 
$V_{\rm eff} (r)$ arising from this reduction 
framework~\cite{jensen_Phys_Rep_Efimov},
\begin{equation}
V_{\rm eff} (r)
=
-
\frac{ g_{\rm eff} }{r^{2}}
\, ,
\;  \; \; \; \; 
\lambda_{\rm eff}
= 
(d-1)^{2}
 + \Theta_{\rm eff}^{2}
\, ,
\;  \; \; \; \; 
d_{\rm eff}=2d
\;  ,
\label{eq:Efimov_effective_potentials}
\end{equation}
where $g_{\rm eff}$
and $\lambda_{\rm eff}$ are related as described
after Eq.~(\ref{eq:ISP_Hamiltonian_unregularized_anisotropic}).

Incidentally, due to the interdimensional equivalence of 
Appendix~\ref{sec:dimensionalities_and_interdimensional},
this result is often quoted in its one-dimensional reduced form
[from Eq.~(\ref{eq:reduction_to_1D})],
$\lambda (d=1)
= \lambda_{\rm eff} - 
(d-1)^{2} + 1/4
= 
  \Theta_{\rm eff}^{2}  + 1/4
$.
For example,
 for the all-important case of ordinary three-dimensional space,
$d_{\rm eff}=6$ and
$\lambda_{\rm eff} = 4+   \Theta_{\rm eff}^{2} $.
Furthermore,
the coupling constant in Eq.~(\ref{eq:Efimov_effective_potentials})
depends upon the physical parameters
defining the system:
when the scattering lengths are large,
it is function of the three ratios of particle masses.
In particular, for the lowest angular eigenvalue of a
three-body three-dimensional
system of identical bosons with zero-range two-particle 
interactions,
the characteristic conformal parameter~(\ref{eq:Theta_ISP_parameter})
is approximately given by the solution of the transcendental 
equation~\cite{jensen_Phys_Rep_Efimov}
\begin{equation}
8 \,
\sinh \left( \frac{\pi \Theta_{\rm eff} }{6} \right)
=
\sqrt{3} 
\,
\Theta_{\rm eff}
\cosh
\left(
\frac{\pi \Theta_{\rm eff}}{2}
\right)
\; ,
\end{equation}
so that $\Theta_{\rm eff} \approx 1.006$,
which corresponds to the strong-coupling regime.

In short, the essential feature shared by the problems discussed above
is the existence of an {\em effective\/} description in terms of  
SO(2,1) conformal invariance,
which results from a prescribed {\em reduction\/} framework.
We now turn our attention to this generic effective problem,
characterized by the Hamiltonian
of Eq.~(\ref{eq:ISP_Hamiltonian_unregularized}).
As different dimensionalities are required for 
the applications to which Eq.~(\ref{eq:ISP_Hamiltonian_unregularized}) refers,
we will  analyze this problem for the arbitrary $d$-dimensional case.
Our goal is to investigate and characterize the possible realization of a conformal anomaly
within this scope.

\section{Conformal Anomaly and Short-Distance Physics}
\label{sec:CA_UV_physics}

Conformal 
symmetry is 
guaranteed at the quantum level when  
the naive scaling of operators, described by the
algebra~(\ref{eq:naive_commutators}), is maintained.
A measure of the deviation from this scaling is afforded by the
 ``anomaly''~\cite{camblong:anom_delta}
\begin{eqnarray}
{\mathcal A} ({\bf r})
& \equiv &
\frac{1}{i \hbar}
[D,H] +  H
=
\left[
\openone
+
\frac{1}{2}
\,
{\mathcal E}_{\bf r}
 \right]
V ({\bf r})
\label{eq:time_rate_of_dilation_op}
\\
& =& 
\frac{r^{d-2}}{2}
\;
{\bf \nabla}
\!
\cdot
\!
\left[
\frac{ {\bf r} \,
 V ({\bf r})}{r^{d-2}} 
\right]
\;  
\label{eq:time_rate_of_dilation_op_ddim}
\end{eqnarray}
(valid for arbitrary $d$ spatial dimensions),  in which  $\openone$
is the identity operator and
 ${\mathcal E}_{\bf r} = {\bf r} \cdot 
{\bf \nabla} $.
At first sight,
the right-hand side of
Eq.~(\ref{eq:time_rate_of_dilation_op})
appears to be zero for any scale-invariant potential; however,
upon closer examination,
this apparent cancellation may break down at 
$r=0$,
where the interaction is singular.
Equations~(\ref{eq:time_rate_of_dilation_op}) and (\ref{eq:time_rate_of_dilation_op_ddim})
can be directly applied to any of the interactions within the conformal
quantum mechanics class, and reduce to the familiar results
known for the two-dimensional contact 
interaction~\cite{camblong:anom_delta,esteve:anom_delta}.
However, the most interesting case is provided by the 
Hamiltonian~(\ref{eq:ISP_Hamiltonian_unregularized}),
whose symmetry breaking can be made apparent 
by means of the
formal $d$-dimensional identity
$
{\bf \nabla}
\!
\cdot
\!
\left[
\hat{\bf r}/
r^{d-1} 
\right]
=
\Omega_{d-1}
\,
 \delta^{(d)} ({\bf r}) $,
in which $\Omega_{d-1}$ is the 
surface area of the unit $(d-1)$-sphere $S^{d-1}$;
then,
\begin{equation}
{\mathcal A} ({\bf r})=
- g \,
\frac{ \Omega_{d-1}}{2}
\,
r^{d-2}
\,
\delta^{(d)} ({\bf r})
\;  .
\label{eq:time_rate_of_dilation_op_ddim_ISP}
\end{equation}
Despite its misleading appearance,
this term is {\em not\/} identically equal to zero,
 due to the singular nature of the interaction at $r=0$.
The recognition of this remarkable singular term,
as well as of its regularized and renormalized 
counterparts, leads to the central result of our paper:
the proof of the existence of a conformal anomaly.

However, two important points should be clarified.
First, 
Eq.~(\ref{eq:time_rate_of_dilation_op_ddim_ISP}) is merely a formal identity,
whose physical meaning can only be manifested through appropriate 
integral expressions.
Second,  the coordinate singularity
 highlights the need to
determine the behavior of the wave function near $r=0$.
Therefore,  nontrivial consequences of
Eq.~(\ref{eq:time_rate_of_dilation_op_ddim_ISP})
can  only be displayed by the expectation value with a
normalized state $\left| \Psi \right\rangle$,
\begin{equation}
\frac{d}{dt}
\left\langle
D
\right\rangle_{\scriptstyle \!  \Psi}
=
\left\langle
{\mathcal A} ({\bf r})
\right\rangle_{\scriptstyle \!  \Psi}
=
- g \, 
\frac{ \Omega_{d-1}}{2}
\,
\int
d^{d} {\bf r}
\,
\delta^{(d)} ({\bf r})
\left| r^{\nu} \Psi ({\bf r})
\right|^{2}
\;   .
\label{eq:time_rate_of_dilation_op_ddim-EV_ISP}
\end{equation}
A similar analysis applies to the
anisotropic interaction of Eq.~(\ref{eq:ISP_Hamiltonian_unregularized_anisotropic});
in this case,
\begin{equation}
\frac{d}{dt}
\left\langle
D
\right\rangle_{\scriptstyle \!  \Psi}
=
\left\langle
{\mathcal A} ({\bf r})
\right\rangle_{\scriptstyle \!  \Psi}
  = 
- g \, 
\frac{ \Omega_{d-1}}{2}
\,
\int
d^{d} {\bf r}
\,
\delta^{(d)} ({\bf r})
\left| r^{\nu} \Psi ({\bf r})
\right|^{2}
\;
F({\bf \Omega})
\;   .
\label{eq:time_rate_of_dilation_op_ddim-EV_ISP_anisotropic}
\end{equation}

It should be noticed that the intermediate steps leading to 
Eqs.~(\ref{eq:time_rate_of_dilation_op_ddim-EV_ISP}) 
and (\ref{eq:time_rate_of_dilation_op_ddim-EV_ISP_anisotropic}) 
are based on formal identities
involving the $d$-dimensional $\delta$ function. 
For the unregularized inverse square potential,
the integrals in Eqs.~(\ref{eq:time_rate_of_dilation_op_ddim-EV_ISP}) 
and (\ref{eq:time_rate_of_dilation_op_ddim-EV_ISP_anisotropic}) 
select the limit $r \rightarrow 0$ 
of the product $r^{\nu  } \Psi ({\bf r})$, which is known to be
proportional to
a Bessel function of  order $i \Theta$, with
$\Theta $ defined
in Eq.~(\ref{eq:Theta_ISP_parameter}).
This limit is ill defined in the strong-coupling regime,
due to the uncontrolled oscillatory behavior of
the Bessel functions of imaginary order.
Consequently, a regularization procedure is called for;
{\em inter alia\/},
this procedure 
will assign a meaningful value to Eqs.~(\ref{eq:time_rate_of_dilation_op_ddim-EV_ISP})
and (\ref{eq:time_rate_of_dilation_op_ddim-EV_ISP_anisotropic}).

\section{Regularization and Renormalization: The Effective-Field Theory 
Program}
\label{sec:regularization_renormalization}

The Hamiltonian~(\ref{eq:ISP_Hamiltonian_unregularized}),
in the strong-coupling regime,
describes an effective system with
singular behavior for short-distance scales.
This interpretation, in which
regularization and renormalization are mandatory,
is inspired  by the {\em effective-field theory  program\/}~\cite{EFT}.
The required regularization procedure is implemented in real space,
where the ultraviolet physics is replaced over length scales $r \alt a$.
The effective theory that comes out of this renormalization is
expected to be applicable within a domain of energies
of magnitude $|E| 
\ll
E_{a} 
\equiv \hbar^{2}/2m a^{2} $.
The scale
$E_{a} $ 
defines an approximate limit of the conformal regime from the
ultraviolet side; effectively, this limit prevents the singular interaction 
from yielding unphysical divergent results for supercritical coupling.

Specifically,
 we consider
a generic class of regularization schemes that explicitly modify the 
ultraviolet physics; each scheme is described by a potential $V^{(<)}({\bf r})  $,
for $ r \alt a$, where $a$ is a small real-space regulator.
An appropriate procedure for the  selection of 
solutions of this singular conformal interaction 
was proposed in Ref.~\cite{landau:77},
using a constant potential for
$ r \alt a$.
Our approach is based on a 
generalization of this method, in which
 a core interaction  
 $V^{(<)}({\bf r})$
is introduced.

Incidentally, in this section, we consider a core
 $V^{(<)}({\bf r}) \equiv V^{(<)}( r) $
with central symmetry $V^{(<)}( r) $. Even though this condition is
not strictly necessary, it leads to a tractable derivation.
Moreover, it is also
consistent with the original rotational invariance of the 
isotropic singular interaction
and  captures the singular behavior 
of the potential, which originates from its radial dependence
(even in the anisotropic case).
The generalization for an anisotropic 
conformal interaction is nontrivial, but when this interaction is reduced to
an effective radial problem, the procedure developed in this section
can be applied.  

The core interaction
is subject to the conditions 
of finiteness 
\begin{equation}
-\infty <
V_{0} \equiv {\rm min} \left[  V^{(<)}(r) \right]  
\equiv - \frac{\hbar^{2}}{2m} \, \frac{\aleph}{a^{2}} 
\; 
\label{eq:core_interaction_restriction:continuity}
\end{equation}
and continuous matching 
with the external inverse square potential at $r=a$, 
\begin{equation}
  V^{(<)}(a) =    V^{(>)}(a) =  -g/a^{2}
 \; .
\label{eq:core_interaction_restriction:matching}
\end{equation}
It should be noticed that
these restrictions imply that $V_{0}<0$ or
$\aleph >0$,
and that 
$\aleph =
\lambda + \varsigma$,
where $\varsigma >0$ is the dimensionless energy difference between 
the minimum $V_{0}$ and the matching 
value~(\ref{eq:core_interaction_restriction:matching}).
In addition, in this approach,
the energies for the interior problem
will be conveniently redefined from the minimum value $V_{0}$;  specifically,
\begin{equation}
U(r) \equiv V^{(<)}(r) -V_{0}
\;    ,  
\; \; \; \; \; \; \;    
\tilde{E}= E -V_{0}
\; .
\label{eq:energies_from_bottom}
\end{equation}

For the spherically symmetric long-range conformal 
interaction of Eq.~(\ref{eq:ISP_Hamiltonian_unregularized}),
central symmetry leads to the separable solution
 \begin{equation}
\Psi ({\bf r})= 
\frac{
\check{Y}_{l{\bf m}}( {\bf \Omega})  \,  v (r) 
}{ r^{\nu } }
\;    ,
\; \; \; \; \; \; \; 
\int d \Omega_{d-1} |\check{Y}_{l{\bf m}}( {\bf \Omega})  |^{2} = \Omega_{d-1}
\;  ,
\label{eq:separable_wf-ultraspherical}
\end{equation}
in which  $\check{Y}_{l{\bf m}}( {\bf \Omega})  $ stands for the ultraspherical 
harmonics on $S^{d-1}$~\cite{ultraspherical_harmonics}, which have been  
conveniently redefined with a normalization integral equal to the solid 
angle $\Omega_{d-1}$. Then, the
 corresponding effective radial Schr\"odinger equation 
for $v(r)$ becomes
\begin{equation}
\left\{
\frac{d^{2}}{dr^{2}}
+
\frac{1}{r} 
\,
\frac{d}{dr}
+
\left[ k^{2} - \frac{(l+\nu)^{2}}{ r^{2} } - {\mathcal V} (r)
 \right]
\right\} v (r)
= 0
\; ,
\label{eq:reduced_Schrodinger_v}
\end{equation}
where ${\mathcal V} (r) = 2 m  V (r)/\hbar^{2}$ and
 $k^{2} = 2 m  E/\hbar^{2}$.
In particular,
for bound states, $k= i \kappa$
and Eq.~(\ref{eq:reduced_Schrodinger_v})
provides solutions of the form
\begin{equation}
v_{a}(r) 
= \left\{ 
\begin{array} {ll} 
v^{(<)}(r)= B_{l,\nu} \,
w_{l+\nu}( \tilde{k} r; \tilde{k})
  &  \mbox{for
$r \leq a$} \; , 
\\ 
 v^{(>)}(r)= A_{l,\nu} \,
K_{i \Theta } (\kappa r)
 &  \mbox{for $r\geq a$ } 
\; ,
\end{array}
\right.
\label{eq:wf_ISP_Landau_regularized}
\end{equation}
in which 
$K_{i \Theta } (z)$ 
is the Macdonald function~\cite{macdonald_function},
and where $\tilde{k}$ is defined from
$\tilde{E}= \hbar^{2} \tilde{k}^{2}/2m$, so that
$\tilde{k}=\sqrt{ -\kappa^{2} - {\mathcal V}_{0}} 
$, with  ${\mathcal V}_{0} = 2 m  V_{0}/\hbar^{2} <0$.
The regularizing core is arbitrary
and  $w_{l+\nu}( \tilde{k} r; \tilde{k})$ is a particular real solution
in that region,
\begin{equation}
\left\{
\frac{d^{2}}{dr^{2}}
+
\frac{1}{r} 
\,
\frac{d}{dr}
+
\left[ \tilde{k}^{2} - \frac{(l+\nu)^{2}}{ r^{2} } - {\mathcal U} (r)
 \right]
\right\}
 w_{l+\nu}( \tilde{k} r; \tilde{k})
= 0
\; ,
\label{eq:reduced_Schrodinger_v_interior}
\end{equation}
where $ {\mathcal U} (r) =  {\mathcal V} (r) -  {\mathcal V}_{0}$;
as an example, $w_{l+\nu}( \tilde{k} r; \tilde{k})$ is
  a Bessel function of order $l+\nu$ when the potential
$V^{(<)}({\bf r})  $ is a constant.

The solution~(\ref{eq:wf_ISP_Landau_regularized}) can be completely
determined by enforcing the following three additional physical conditions:
(a) continuity at  $r=a$
of the radial wave function;
(b)
continuity at  $r=a$ of the logarithmic derivative
of the radial wave function;
and (c) normalization of the wave function. 
In what follows, these conditions will be stated 
using the auxiliary  parameters 
\begin{equation}
\xi
 = 
\kappa a
\, ,
\; \; \; \;  \; \; \; \; 
\tilde{\xi}
 = 
\tilde{k} a
\, ,
\label{eq:dimensionless_wave_number_definitions}
\end{equation}
which satisfy Eq.~(\ref{eq:energies_from_bottom}), i.e.,
\begin{equation}
\tilde{\xi}^2
+ \xi^{2} 
= 
 \aleph
\;  .
\label{eq:pythagorean}
\end{equation}
Consequently, these conditions (a)-(c)
become, respectively,
\begin{equation}
 B_{l,\nu} \, 
w_{l+\nu}( \tilde{\xi}; \tilde{k})
= A_{l,\nu} \, K_{i \Theta}  ( \xi)
\;  ,
\label{eq:continutity_WF}
\end{equation}
\begin{equation}
{\cal L}_{l+\nu}^{(<)} (\tilde{\xi}; \tilde{k})
=
{\cal L}_{i \Theta}^{(>)} (\xi)
\; ,
\label{eq:continutity_log_derivatives_WF}
\end{equation}
and [cf. Eq.~(\ref{eq:separable_wf-ultraspherical})]
\begin{equation}
\int
d^{d} {\bf r} \,
| \Psi ({\bf r}) |^{2}
=
\Omega_{d-1} \,
\int_{0}^{\infty} 
 dr r | v(r) |^{2}
=
1
\; ,
\label{eq:normalization_WF}
\end{equation}
where we have conveniently redefined
the logarithmic derivatives from
${\cal L}_{i \Theta}^{(>)} (\xi) \equiv 
{\mathcal E}_{\xi} 
\left[
\ln K_{i \Theta} (\xi) \right]$, with
${\mathcal E}_{\xi} = \xi \partial/\partial \xi$,
and similarly for 
$
{\cal L}_{l+\nu}^{(<)} (\tilde{\xi}; \tilde{k})
$ 
in terms of
$w_{l+\nu}( \tilde{\xi}; \tilde{k})$.
Explicitly, Eq.~(\ref{eq:normalization_WF}) takes the form
\begin{equation}
 \Omega_{d-1} 
\,
\left[
B_{l,\nu}^{2}
\,
\tilde{k}^{-2}
\,
{\mathcal J}_{l+\nu}(\tilde{\xi}; \tilde{k} ) 
\!
+
\!
A_{l,\nu}^{2}
\,
\kappa^{-2}
\,
\!
{\mathcal K}_{i \Theta }(\xi)
\right]
=
1
\; ,
\label{eq:normalization_WF_explicit}
\end{equation}
in which the normalization constants can be chosen to be real, and
where
\begin{equation}
{\mathcal K_{i \Theta}} (\xi)
 \equiv
\int_{\xi}^{\infty} 
d z \, z
 \left[ K_{i \Theta} (z) \right]^{2} 
\label{eq:mathcal_K_integral_def}
\end{equation}
and
\begin{equation}
{\mathcal J_{l+\nu}} ( \tilde{\xi};  \tilde{k}  )
  \equiv  
\int_{0}^{  \tilde{\xi}  } 
d z \, z
 \left[ 
w_{l+\nu}( z; \tilde{k})
\right]^{2} 
\;  .
\label{eq:mathcal_J_integral_def}
\end{equation}
Equations~(\ref{eq:continutity_WF}) and (\ref{eq:normalization_WF_explicit})
then provide the values of the constants  $ A_{l,\nu} $ and $ B_{l,\nu}$;
for example, 
\begin{equation}
B_{l,\nu}
=
\frac{ \kappa}{ \sqrt{ \Omega_{d-1} } }
\,
\left\{
\frac{ \xi^{2} }{ \tilde{\xi}^{2} } 
\,
{\mathcal J}_{l+\nu}(\tilde{\xi}; \tilde{k} ) 
\!
+
\!
\left[
\frac{
 w_{l+\nu}  ( \tilde{\xi} ; \tilde{k}) }{  K_{i \Theta}  (\xi) }
\right]^{2} \!
\!
{\mathcal K}_{i \Theta }(\xi)
\right\}^{-1/2}
\; .
\label{eq:coeff_B_value}
\end{equation}
For reasons that will become
clear in the next section, it is convenient to rewrite 
Eqs.~(\ref{eq:mathcal_K_integral_def}) and (\ref{eq:mathcal_J_integral_def})
in an alternative way, using the generalized Lommel integrals of 
Appendix~\ref{sec:generalized_Lommel_integrals}.
First, 
the integral defined by
Eq.~(\ref{eq:mathcal_K_integral_def}),
which applies to the external domain ($r \geq a$), 
can be expressed as
\begin{equation}
{\mathcal K_{i \Theta}} (\xi)
=
\frac{ 1}{2}
\left[K_{i \Theta} (\xi) \right]^{2}
\,
{\mathcal M}^{(>)}_{i \Theta} ( \xi ) 
\; ,
\label{eq:Lommel_for_mathcal_K_integral}
\end{equation}
where
\begin{equation}
{\mathcal M}_{i\Theta}^{(>)} (\xi) 
\equiv 
\left[ {\mathcal L}^{(>)}_{i \Theta} (\xi)     \right]^{2} 
+ \Theta^{2} - \xi^{2}
\; .
\label{eq:matching_function_>}
\end{equation}
Similarly, 
the integral defined by
Eq.~(\ref{eq:mathcal_J_integral_def}),
which applies to the internal domain ($r \leq a$), 
 takes the form
\begin{equation}
{\mathcal J_{l+\nu}} ( \tilde{\xi};  \tilde{k}  )
=
\frac{1}{2}  \left[ w_{l+\nu} (\tilde{\xi};  \tilde{k} ) \right]^{2} 
{\mathcal M}^{(<)}_{l+\nu} ( \tilde{\xi};  \tilde{k}  ) 
+
{\mathcal U}_{l+\nu} ( \tilde{\xi};  \tilde{k}  ) 
\; ,
\label{eq:Lommel_for_mathcal_J_integral}
\end{equation}
where
\begin{equation}
{\mathcal M}^{(<)}_{l+\nu} ( \tilde{\xi};  \tilde{k}  ) 
\!
 \equiv 
\!
\left[
 {\mathcal L}^{(<)}_{l+\nu}  (\tilde{\xi}; \tilde{k}) \right]^{2} 
+ \left[
\tilde{\xi}^{2} - 
\!
(l+\nu)^{2} 
-
\tilde{\xi}^{2} 
\check{U} (\tilde{\xi}; \tilde{k})
\right]
\;  
\label{eq:matching_function_<}
\end{equation}
and
\begin{equation}
{\mathcal U}_{l+\nu} ( \tilde{\xi};  \tilde{k}  ) 
\!
 \equiv 
\!
\int_{0}^{  \tilde{\xi}  } 
d z \, z
 \left[ w_{l+\nu} (z;  \tilde{k} ) \right]^{2} 
\!
\left[
\!
\left(
\openone
+
\frac{1}{2}
\,
{\mathcal E}_{z}
 \right)
\!
\check{U} (z; \tilde{k})
\right]
,
\label{eq:mathcal_U_integral_def}
\end{equation}
with $\check{U} \equiv U/\tilde{E}$ and 
${\mathcal E}_{z} = z \partial/\partial z$.
The Lommel integral relation~(\ref{eq:Lommel_for_mathcal_K_integral})
appears to be simpler than Eq.~(\ref{eq:Lommel_for_mathcal_J_integral})
because of the absence of an extra core 
$\check{U} (z; \tilde{k})$ in the external domain.

 In addition, the continuity conditions
of the potential, Eq.~(\ref{eq:core_interaction_restriction:matching}),
and of the logarithmic derivatives,
Eq.~(\ref{eq:continutity_log_derivatives_WF}),
imply the equality of the ``matching functions''~(\ref{eq:matching_function_>})
and (\ref{eq:matching_function_<}), i.e.,
\begin{equation}
{\mathcal M}^{(<)}_{l+\nu} ( \tilde{\xi};  \tilde{k}  ) 
=
{\mathcal M}^{(>)}_{i \Theta} ( \xi ) 
\; .
\label{eq:matching_functions}
\end{equation}
As a corollary,  a combined Lommel relation can be obtained by elimination
of the matching functions from 
Eqs.~(\ref{eq:Lommel_for_mathcal_K_integral})
and (\ref{eq:Lommel_for_mathcal_J_integral}),
\begin{equation}
{\mathcal J_{l+\nu}} ( \tilde{\xi};  \tilde{k}  )
-
{\mathcal U}_{l+\nu} ( \tilde{\xi};  \tilde{k}  ) 
=
\left[
\frac{
 w_{l+\nu}  ( \tilde{\xi} ; \tilde{k}) }{  K_{i \Theta}  (\xi) }
\right]^{2} \!
\!
{\mathcal K}_{i \Theta }(\xi)
\; .
\label{eq:combined_Lommel}
\end{equation}

Even though the 
implementation of a renormalization procedure 
is a necessary condition for 
the emergence
of the conformal anomaly,
 the actual details of this procedure are not 
explicitly required. 
It suffices to know that
these details are to be consistently derived from 
Eqs.~(\ref{eq:dimensionless_wave_number_definitions})--(\ref{eq:combined_Lommel}),
which permit the exact evaluation of all relevant expectation values,
and
 by enforcing the finiteness of
a  particular bound state energy.

\section{Computation of the Conformal Anomaly}
\label{sec:CA_computation}

The value of the anomalous part
of the commutator
$[D,H]$ is given  as the ``anomaly''
${\mathcal A} ({\bf r})$ in Eq.~(\ref{eq:time_rate_of_dilation_op}).
In Sec.~\ref{sec:CA_UV_physics},
this quantity was computed for the unregularized inverse square potential
in terms of the formal identity~(\ref{eq:time_rate_of_dilation_op_ddim_ISP});
this expression, in turn, led to an ill-defined 
expectation value~(\ref{eq:time_rate_of_dilation_op_ddim-EV_ISP}).
This difficulty can be overcome
when the singular conformal interaction is regularized according to the
generic scheme introduced in Sec.~\ref{sec:regularization_renormalization}.
Then, 
Eq.~(\ref{eq:time_rate_of_dilation_op})
will in principle yield two different contributions:
one for $r \leq a$ and one for $r \geq a$, with the latter being 
of the form~(\ref{eq:time_rate_of_dilation_op_ddim_ISP}); thus,
\begin{equation}
{\mathcal A}_{a} ({\bf r})
=
\left[
\left(
\openone
+
\frac{1}{2}
\,
{\mathcal E}_{\bf r}
 \right)
\,
V^{(<)} ({\bf r})
\right]
 \,
\theta (a-r) 
- g \,
\frac{ \Omega_{d-1}}{2}
\,
r^{d-2}
\,
\delta^{(d)} ({\bf r})
 \,
 \theta (r-a) 
\;  ,
\label{eq:time_rate_of_dilation_op_ddim_ISP_reg}
\end{equation}
where  $\theta (z)$ stands for the Heaviside function,
is the regularized 
counterpart of Eq.~(\ref{eq:time_rate_of_dilation_op_ddim_ISP}).
Explicitly, this leads to an expectation value
\begin{equation}
\frac{d}{dt}
\left\langle
D
\right\rangle_{\scriptstyle \!  \Psi}
=
\left[
\left\langle
{\mathcal A}_{a}  ({\bf r})
\right\rangle_{\scriptstyle \!  \Psi_{a} }^{(<)}
+
\left\langle
{\mathcal A}_{a}  ({\bf r})
\right\rangle_{\scriptstyle \!  \Psi_{a} }^{(>)}
\right]
\;  ,
\label{eq:time_rate_of_dilation_op_ddim-EV_ISP_reg}
\end{equation}
where the integration range
is split into the two regions: $0 \leq r \leq a$ and $r \geq a$.
Moreover,
 the identically vanishing second term
\begin{equation}
\left\langle
{\mathcal A}_{a} 
({\bf r})
\right\rangle_{\scriptstyle \!  \Psi_{a} }^{(>)}
= 0 
\label{eq:vanishing_external_contribution_to_anomaly}
\end{equation}
in Eq.~(\ref{eq:time_rate_of_dilation_op_ddim_ISP_reg})  shows that the source 
of the conformal anomaly is confined to an arbitrarily small region about 
the origin.
This result can be confirmed from
a  straightforward 
replacement of Eq.~(\ref{eq:time_rate_of_dilation_op}) by
${\mathcal A} ({\bf r})=-\frac{1}{2} ( d - 2 )\, V ({\bf r})
+
\frac{1}{2}
\,
{\bf \nabla}
\!
\cdot
\!
\left\{
{\bf r} \,
 V ({\bf r})
\right\}
$,
which 
is identically equal to zero 
for any domain that excludes the origin, when applied to
 any homogeneous potential of degree -2
(a defining
characteristic of the external conformal interaction).

Once Eq.~(\ref{eq:vanishing_external_contribution_to_anomaly}) is established,
the anomaly can be computed from the contribution arising 
from the ultraviolet domain $r \leq a$,
\begin{equation}
\frac{d}{dt}
\left\langle
     D
\right\rangle_{\scriptstyle \!  \Psi}
=
\left\langle
{\mathcal A}_{a}  ({\bf r})
\right\rangle_{\scriptstyle \!  \Psi_{a} }^{(<)}
  = 
\int_{r \leq a} d^{d} {\bf r} 
\,
\left[
\left(
\openone
+
\frac{1}{2}
\,
{\mathcal E}_{\bf r}
 \right)
V(r)
\right]
\,
|\Psi_{a} ({\bf r})|^{2}
\; .
\label{eq:anomaly_internal_region_0}
\end{equation}
In Eq.~(\ref{eq:anomaly_internal_region_0}),
$V(r) \equiv V^{(<)} (r)$ can be replaced 
using Eq.~(\ref{eq:energies_from_bottom}),
and 
$\Psi_{a} ({\bf r})$ using Eqs.~(\ref{eq:separable_wf-ultraspherical})
and (\ref{eq:wf_ISP_Landau_regularized}); when these 
substitutions are made and  the dimensionless variable
$\tilde{\xi}$ in Eq.~(\ref{eq:dimensionless_wave_number_definitions}) is introduced, 
Eq.~(\ref{eq:anomaly_internal_region_0}) becomes
\begin{equation}
\frac{d}{dt}
\left\langle
     D
\right\rangle_{\scriptstyle \!  \Psi}
=
\frac{\Omega_{d-1}  \, B_{l,\nu}^{2} 
}{\tilde{k}^{2} }
\, \int_{0}^{  \tilde{\xi}  } 
d z \, z
\,
 \left[ 
w_{l+\nu}( z; \tilde{k})
 \right]^{2} 
\,
\left\{
\left(
\openone
+
\frac{1}{2}
\,
{\mathcal E}_{z}
 \right)
\left[  V_{0} + U \left(  \frac{z}{\tilde{k}}  \right)
\right]
\right\}
\,  .
\label{eq:anomaly_internal_region_1}
\end{equation}
Despite its cumbersome appearance,
the integral in Eq.~(\ref{eq:anomaly_internal_region_1}) 
can be easily evaluated once the  
definitions~(\ref{eq:mathcal_J_integral_def})
and
(\ref{eq:mathcal_U_integral_def}) are introduced, so that
\begin{eqnarray}
\frac{1}{E}
\,
\frac{d}{dt}
\left\langle
     D
\right\rangle_{\scriptstyle \!  \Psi}
 &  =  &
\frac{1}{E}
\,
\frac{\Omega_{d-1}  \, B_{l,\nu}^{2} 
}{\tilde{k}^{2} }
\left[
V_{0}
\,
{\mathcal J_{l+\nu}} ( \tilde{\xi};  \tilde{k}  )
+
\tilde{E}
\,
{\mathcal U}_{l+\nu} ( \tilde{\xi};  \tilde{k}  ) 
\right]
\; 
\label{eq:anomaly_internal_region_2}
\\
&  = &
\frac{\Omega_{d-1}  \, B_{l,\nu}^{2}}{ \kappa^{2} } 
\,
\left\{
\frac{ \xi^{2} }{ \tilde{\xi}^{2} } 
\,
{\mathcal J}_{l+\nu}(\tilde{\xi}; \tilde{k} ) 
\!
+
\!
\left[
{\mathcal J}_{l+\nu}(\tilde{\xi}; \tilde{k} ) 
-
{\mathcal U}_{l+\nu} ( \tilde{\xi};  \tilde{k}  ) 
\right]
\right\}
\,  ,
\label{eq:anomaly_internal_region_3}
\end{eqnarray}
where 
$V_{0}$ was replaced through the relation~(\ref{eq:energies_from_bottom})
or (\ref{eq:pythagorean}),
and $E=-\hbar^{2} \kappa^{2}/2m$.
Furthermore,
in Eq.~(\ref{eq:anomaly_internal_region_3}),
the difference
$
{\mathcal J}_{l+\nu}(\tilde{\xi}; \tilde{k} ) 
-
{\mathcal U}_{l+\nu} ( \tilde{\xi};  \tilde{k}  ) 
$
can be evaluated employing Eq.~(\ref{eq:combined_Lommel}),
so that
\begin{equation}
\frac{1}{E}
\,
\frac{d}{dt}
\left\langle
     D
\right\rangle_{\scriptstyle \!  \Psi}
=
\frac{\Omega_{d-1}  \, B_{l,\nu}^{2}}{ \kappa^{2} } 
\,
\left\{
\frac{ \xi^{2} }{ \tilde{\xi}^{2} } 
\,
{\mathcal J}_{l+\nu}(\tilde{\xi}; \tilde{k} ) 
\!
+
\!
\left[
\frac{
 w_{l+\nu}  ( \tilde{\xi} ; \tilde{k}) }{  K_{i \Theta}  (\xi) }
\right]^{2} \!
\!
{\mathcal K}_{i \Theta }(\xi)
\right\}
\,  .
\label{eq:anomaly_internal_region_4}
\end{equation}
Finally, the coefficient $B_{l,\nu}$
can be eliminated using
Eq.~(\ref{eq:coeff_B_value}), 
which shows that the right-hand side of 
Eq.~(\ref{eq:anomaly_internal_region_4})
is identically equal to one for {\em any\/} bound state.
This remarkable simplification
concludes the proof that
the anomaly
defined in Eq.~(\ref{eq:time_rate_of_dilation_op})
is indeed given by
\begin{equation}
\frac{d}{dt}
\left\langle
D
\right\rangle_{\scriptstyle \!  \Psi}
= 
E
\;  ,
\label{eq:anomaly_ISP}
\end{equation}
where
$E$ is the energy of the corresponding stationary normalized state.

In short, we have validated the relation~(\ref{eq:anomaly_ISP})---which agrees with the 
{\em formal\/} prediction
from properties of expectation values~\cite{camblong:anom_delta}.
This validation has been established
using  a generic regularization procedure. Therefore, 
regardless of the renormalization framework used, 
an {\em anomaly\/} is generated.
The generality of Eq.~(\ref{eq:anomaly_ISP})
 makes it available for a variety of physical applications,
and is a {\em necessary condition when the theory is renormalized\/}.

\section{Renormalization Frameworks}
\label{sec:renormalization_frameworks}

 In the previous section we showed that the property~(\ref{eq:anomaly_ISP})
and related symmetry-breaking results are
independent of the details of the regularization procedure.
Because of the {\em generality\/} of the real-space regularization approach
presented in this paper,
these results extend the two-dimensional 
analysis of Ref.~\cite{cam:anomaly_ISP_2D} in a number of nontrivial ways:

(i) For arbitrary renormalization frameworks, other than the ``intrinsic''
one of Ref.~\cite{cam:anomaly_ISP_2D}
(see below).

(ii) For any dimensionality $d$. Again, the two-dimensional case 
of Ref.~\cite{cam:anomaly_ISP_2D} has unique features that considerably simplify the
derivation within the intrinsic framework.
 This is particularly relevant because the physical
applications that appear to be most interesting are 
$d$-dimensional realizations of this phenomenon,
with $d \equiv d_{\rm eff} \neq 2$.

(iii) For any bound state and angular momentum channel
(and not just for the $l=0$ channel associated with the ground state considered in 
Refs.~\cite{camblong:anom_delta,esteve:anom_delta,cam:anomaly_ISP_2D}).

In this section we highlight
the relevance of these results with
an overview of the
real-space ``effective,'' ``intrinsic,'' and ``core'' renormalization
frameworks 
(according to the presentation of Ref.~\cite{renormalization_CQM}),
and discuss their relationship to the present anomaly calculation.
Despite their apparent differences, 
these frameworks share the basic physical requirement that the system 
is renormalized
under the assumption that the ultraviolet physics
dictates the possible existence of bound states of finite energy;
the corresponding energies $E$ and 
values of $\kappa \propto \sqrt{ |E|} $
are then required to remain finite.

In order to facilitate the implementation of this renormalization program,
it is convenient to display the specific limiting form that
Eq.~(\ref{eq:continutity_log_derivatives_WF})
takes when 
$a \rightarrow 0$; more precisely,
\begin{equation}
\cot \left[
\alpha ,
\left( 
\Theta, \kappa a 
\right)
\right] 
  \stackrel{( \kappa a \ll 1)}{\sim}
\frac{1}{\Theta}
 \;
{\cal L}^{(<)} (\aleph)
\;  ,
\label{eq:log_derivatives_asymptotic}
\end{equation}
where
${\cal L}^{(<)} (\tilde{k} 
a; \tilde{k} )
  \stackrel{( \kappa a \ll 1)}{\sim}
{\cal L}^{(<)}  (\aleph)$
and
\begin{equation}
\alpha \left( \Theta, \kappa a \right)
\equiv
\Theta \left[ \ln \left( \frac{\kappa a}{2} \right) + \gamma_{\Theta} 
\right]
\;  ,
\label{eq:alpha}
\end{equation}
with $\gamma_{\Theta} = - 
\left\{
{\rm phase} 
\left[
\Gamma (1 +i\Theta )
\right]
\right\}/\Theta$
(which reduces to the Euler-Mascheroni constant $\gamma$~\cite{euler_mascheroni}
 in the limit $\Theta 
\rightarrow 0$). 

In the
effective renormalization framework, 
the system is regularized maintaining finite values of 
$|E| \ll
E_{a} \equiv \hbar^{2}/2m a^{2} $.
This condition
defines an asymptotic conformally invariant domain;
within that domain,
the condition
$ 
\kappa a \ll 1
$
limits the ultraviolet applicability of this effective scheme.
Most importantly, this condition is
systematically applied to derive physical predictions in  a direct manner,
within the prescriptions of Sec.~\ref{sec:regularization_renormalization}.
As a result, Eq.~(\ref{eq:log_derivatives_asymptotic})
leads to the
bound-state energy levels~\cite{renormalization_CQM} 
\begin{equation}
E_{ n }
=
E_{0}
\,
\exp \left( - \frac{2 \pi n}{\Theta } \right)
\;  ,
\label{eq:cutoff_BS_regularized_energies_phenomenological}
\end{equation}
in which  $E_{0}<0$ is an arbitrary 
proportionality constant.
This derivation also shows that, as
ultraviolet physics sets in for
$|E| \agt
E_{a} $
(that is, for $\kappa a \agt 1$),
no claim can be made as to 
the nature of the states on these deeper scales.

A few comments are in order regarding
Eq.~(\ref{eq:cutoff_BS_regularized_energies_phenomenological}).
First, it explicitly displays a breakdown of the conformal symmetry,
by the introduction
of a scale $|E_{0}|$ and an associated sequence of bound states.
Second,
the scale $|E_{0}|$ arises from
the renormalization procedure.
Third, 
as a renormalization scale,
$|E_{0}|$
cannot be predicted by the conformal model
and it is to be adjusted experimentally.
Fourth, 
once the experimental determination is carried out,
an unambiguous prediction 
[from 
Eqs.~(\ref{eq:anomaly_ISP}) 
and (\ref{eq:cutoff_BS_regularized_energies_phenomenological})]
follows,
\begin{equation}
\frac{E_{n+1}}{E_{n}} =
\frac{
\mbox{ {\large $ 
\frac{
\; \; 
d
\left\langle
D
\right\rangle_{\scriptscriptstyle \!  \Psi_{n+1}}
}{dt \; \; \; \; \; \;
}
$}}
}{
\mbox{ {\large $ 
\frac{\; \; \;
d
\left\langle
D
\right\rangle_{\scriptscriptstyle \!  \Psi_{n}} \; }{dt \; \; \;}
$}}
}
= 
\exp \left( - \frac{2 \pi }{\Theta } \right)
\;  ,
\label{eq:anomaly_ISP_phenomenological}
\end{equation}
within the range of applicability,
$ 
\kappa a \ll 1
$.
This is in agreement with the conclusions of phenomenological analyses
of the Efimov effect~\cite{jensen_Phys_Rep_Efimov}.

The alternative 
intrinsic and core frameworks are characterized by the fact that
the limit $\xi = \kappa a \rightarrow 0$
is {\em strictly\/} applied before drawing any conclusions
about the physics. Therefore,
in order to 
keep the bound-state energies and $\kappa$ values finite,
a {\em running\/} coupling parameter is explicitly introduced,
so that Eq.~(\ref{eq:log_derivatives_asymptotic}) 
is still maintained in this limit. 
The running parameter is either the 
conformal coupling
$g$, in the intrinsic framework, 
or the strength $\aleph$ 
of the 
regularizing core interaction, in the core framework.

In the case of the intrinsic framework,
the dependence $g=g(a)$, equivalent to
$ \Theta = \Theta (a)$, is enforced.
This leads to the asymptotic running behavior
 $\Theta \sim 0$, which ensures that
the  left-hand side of Eq.~(\ref{eq:log_derivatives_asymptotic})
remains well defined.
This limiting procedure 
leads to the renormalization framework of
Refs.~\cite{gup:93,camblong:isp-dt};
in particular,
Eq.~(\ref{eq:anomaly_ISP_phenomenological})
[with the condition $\Theta \sim 0$]
implies the existence of a single bound state.
In its original form,
the renormalization framework of
Refs.~\cite{gup:93,camblong:isp-dt}
was based upon
a Dirichlet boundary condition,
which we now reinterpret 
as an {\em effective\/}
Dirichlet boundary condition~\cite{pi_collective} 
$
u (r =a)  \stackrel{(a \rightarrow 0)}{\sim}
0$,
for the
reduced radial wave function
$u(r) = \sqrt{r} \,
v(r)$.
This result is
guaranteed by the prefactor $\sqrt{r}$,
regardless of the behavior of $v(r)$. As for $v(r)$,
two distinct cases should be considered:
(i)
the special case characterized by
the simultaneous assignments $d=2$, $l=0$, and constant $V^{(<)} ({\bf r})$
[or, to be more precise, 
with $\varsigma=  | {\mathcal V_{0}} | a^{2}
- \lambda = o (a^{2})$],
for which 
$
\tilde{\xi} = \tilde{k} a = O(\Theta)
$ and
$\cos \alpha (\Theta, \kappa a) 
  \stackrel{(a \rightarrow 0)}{\sim} 
0  
$;
(ii) the generic case, characterized by
$d\neq 2$, 
or $l \neq 0$, or $V^{(<)} ({\bf r})$ not being
constant, for which 
the variable $\tilde{\xi} = \tilde{k} a$
acquires a {\em nonvanishing\/} limit value
$\sqrt{\aleph}= \sqrt{(l+\nu)^{2} + \varsigma} $
[as either 
$(l+ \nu) \neq 0$ or  $\varsigma \neq 0$],
and
$\sin \alpha (\Theta, \kappa a) 
 \stackrel{(a \rightarrow 0)}{\sim} 
0
$.

The smallness of the variable 
$\tilde{\xi}$ 
for case (i) above is the
main reason for the simplicity of the derivation
 of Ref.~\cite{cam:anomaly_ISP_2D}.
In effect, in this case,
Eq.~(\ref{eq:anomaly_internal_region_0}) 
can be approximated 
using the small-argument behavior of Bessel 
functions without 
explicitly computing full-fledged Lommel integrals. Thus,
$  {\mathcal J_{l+ \nu=0}}  (\tilde{k} a )  
 \stackrel{(a \rightarrow 0)}{\sim} 
\Theta^{2}/2
$
and 
$|B_{l=0,\nu=0} |
= |A_{l=0,\nu=0} K_{i\Theta} (\kappa a)/J_{0}(\tilde{k} a) |
 \stackrel{(a \rightarrow 0)}{\sim} 
 \kappa/(\sqrt{\pi}\Theta)$,
leading to
$
d
\left\langle
D
\right\rangle_{\scriptstyle \!  \Psi}/dt
 \stackrel{(a \rightarrow 0)}{\sim} 
 2 \pi B_{0,0}^{2} 
  {\mathcal J_{
0}}  (\tilde{k} a ) V_{0}/\tilde{k}^{2}
 \stackrel{(a \rightarrow 0)}{\sim}  E$,
as discussed in Ref.~\cite{cam:anomaly_ISP_2D}.
By contrast, for the generic case (ii), the analysis presented in this 
paper, based on the theory developed
in Sec.~\ref{sec:regularization_renormalization}
and Appendix~\ref{sec:generalized_Lommel_integrals},
is inescapable.

Finally, in the core renormalization framework,
the strength of the core interaction
becomes a running coupling parameter: $ \aleph = \aleph (a) $,
but 
 the conformal coupling $g$ 
remains constant~\cite{beane:00,bawin:03}.
As a result,
Eq.~(\ref{eq:log_derivatives_asymptotic}) 
provides  
the limit-cycle running
that has been used
in renormalization analyses of the three-body 
problem~\cite{3_body_nucleon,beane:00,bawin:03}.

Incidentally, the ``effective'' renormalization framework 
discussed in Ref.~\cite{renormalization_CQM}
(and summarized in this section)
leads directly to a characterization of the thermodynamics 
of black holes.
In essence, this amounts to a reinterpretation of 
't Hooft's brick wall method~\cite{thooft:85},
in which ultraviolet ``new'' physics 
sets in within a distance of the order of the Planck scale 
from the horizon.
The computation of Appendix~\ref{sec:NH_conformal_behavior}
shows that the leading behavior near the horizon is conformal
and nontrivial, in that the effective system is placed in the 
supercritical regime.
This asymptotic leading contribution, governed by the effective conformal 
interaction, requires renormalization and 
provides the correct thermodynamics~\cite{BH_thermo}.
It should be noticed that there is an  {\em alternative\/} 
treatment, based upon the 
method of self-adjoint extensions,
which  has been recently discussed in 
Refs.~\cite{gov:BH_states,gupta:BH}.

\section{Conclusions}
\label{sec:conclusions}

Realizations of the conformal anomaly
involve a breakdown of the 
associated SO(2,1) algebra.
In this paper
 we have shown
that the actual emergence and value of the conformal
anomaly rely 
upon the application of a renormalization procedure, but are otherwise
 independent of the details of the ultraviolet physics.
In this sense, the  results derived herein 
are robust and totally general.
As such, they are 
intended to shed light on the physics of 
any system
with
a conformally invariant
domain
for which the short-distance physics dictates the
existence of bound states.

In particular, the 
 dipole-bound anions of molecular physics and the
Efimov effect are 
physical realizations 
of this unusual anomaly.
In addition,
the intriguing near-horizon physics of black 
holes appears to suggest yet another 
example of this ubiquitous phenomenon;
the details of the thermodynamics arising from this conformal
description will be reported elsewhere.

\acknowledgments{
This research was supported in part by
NSF under Grant No.\ PHY-0308300 (H.E.C. and C.R.O.) and by
the University of San Francisco Faculty Development Fund
(H.E.C.).
We also thank Dean Stanley Nel for generous travel support
and
Professors 
Cliff Burgess,
Luis N. Epele,  Huner Fanchiotti,
and Carlos A. Garc\'{\i}a Canal
for early discussions of
 this work.}

\appendix

\section{Anisotropic Long-Range Conformal Interaction and
Conformal Behavior of Dipole-Bound Anions}
\label{sec:anisotropic_ISP_and_dipole_bound_anions}

In this appendix we show the mathematical procedure that reduces the
anisotropic inverse square potential
to an {\em effective\/} isotropic 
interaction.

The Schr\"{o}dinger equation for the 
Hamiltonian~(\ref{eq:ISP_Hamiltonian_unregularized_anisotropic})
can be separated in spherical coordinates 
by means of
 \begin{equation}
\Psi ({\bf r})= 
\frac{
\Xi ( {\bf \Omega})  \,  u (r) 
}{ r^{\nu + 1/2 } }
\;    ,
\label{eq:separable_wf-ultraspherical_anisotropic}
\end{equation}
with normalization
 \begin{equation}
\int d \Omega_{d-1} 
\,
| \Xi ( {\bf \Omega})  |^{2} = 
1
\;  .
\label{eq:wf-ultraspherical_anisotropic_normalization}
\end{equation}

As a result,
the angular part 
$\Xi ( {\bf \Omega})  $  of the wave function
is no longer a solution
to Laplace's equation 
on the unit $(d-1)$-sphere $S^{d-1}$;
instead, it satisfies the
modified equation
\begin{equation}
\hat{A} 
\,
\Xi ({\bf \Omega}) 
= \gamma 
\,
\Xi ({\bf \Omega}) 
 \; ,
\label{eq:angular_eq_anisotropic}
\end{equation}
where
\begin{equation}
\hat{A} = - \Lambda^{2} + \lambda 
F ({\bf \Omega}) 
 \; 
\label{eq:angular_operator} 
\end{equation}
 and $\Lambda^{2} =
L^{2} /\hbar^{2}$ is the dimensionless squared angular momentum.
The corresponding radial equation
\begin{equation}
\frac{d^{2}  u(r)}{dr^{2}} + 
\left( k^{2} 
+ \frac{\gamma - \nu^{2} + 1/4}{r^{2}}
\right) u(r) = 0 
\; 
\label{eq:radial_eq_anisotropic}
\end{equation}
is coupled to
Eq.~(\ref{eq:angular_eq_anisotropic}) 
through the
separation constant $\gamma$.
Equation~(\ref{eq:radial_eq_anisotropic})
can be compared against 
the radial equation of an isotropic inverse square potential,
which is obtained by another
 Liouville transformation~\cite{forsyth:Liouville} of the 
form~(\ref{eq:separable_wf-ultraspherical_anisotropic}),
but with ultraspherical harmonics instead of
$\Xi ( {\bf \Omega})  $
and for
$
V_{\rm eff}({\bf r}) \propto r^{-2}$
without angular dependence;
the effective equation
\begin{equation}
\frac{d^{2}  u(r)}{dr^{2}} + 
\left[
 k^{2} 
+ \frac{\lambda_{\rm eff} - (l+ \nu)^{2} + 1/4}{r^{2}}
\right]
 u(r) = 0 
\; 
\label{eq:radial_eq_isotropic}
\end{equation}
is identical to
Eq.~(\ref{eq:radial_eq_anisotropic})
when the following identifications are made:
\begin{equation}
V_{\rm eff}({\bf r}) =
- 
\frac{ g_{\rm eff} 
}{r^{2}}
\, ,
\; \; \; 
g_{\rm eff} 
=
\frac{ \hbar^{2} }{2m}
\lambda_{\rm eff}  
\, ,
\; \; \;
\left. 
\lambda_{\rm eff}  
\right|_{l=0}
=
\gamma 
\; . 
\label{eq:effective_reduction_anisotropic-->>isotropic}
\end{equation}
Consequently,
Eq.~(\ref{eq:radial_eq_anisotropic}) 
can be 
thought of as the radial part of
a $d$-dimensional  {\em effective isotropic\/} 
conformal interaction 
for $l=0$.

 Furthermore, the values $\gamma$ are quantized
from the angular equation~(\ref{eq:angular_eq_anisotropic}) 
and depend upon the coupling $\lambda$ of the anisotropic potential.
This relationship can be made more explicit by expanding,
in the ultraspherical-harmonic basis
$Y_{l {\bf m}}( {\bf \Omega}) $,
the anisotropy factor  
\begin{equation}
 F ({\bf \Omega}) 
=
\sum_{l, {\bf m}} 
F_{l {\bf m}} 
Y_{l {\bf m}}( {\bf \Omega}) 
\;
\end{equation} 
and  the angular wave function
\begin{equation}
\Xi  ({\bf \Omega}) 
=
\sum_{l, {\bf m}} 
\Xi_{l {\bf m}} 
Y_{l {\bf m}}( {\bf \Omega}) 
\; .
\end{equation} 
This decomposition yields the matrix counterpart of
Eq.~(\ref{eq:angular_eq_anisotropic}), whence the
anticipated relationship can be formally displayed 
by the infinite secular determinant
\begin{equation}
D(\gamma, \lambda) \equiv \det M (\gamma,\lambda) 
, 
\; \; \; \; \; \; 
M(\gamma, \lambda) 
= - A(\lambda) + \gamma  \, \openone 
\; ,
\label{eq:det_matrix_angular_eq_anisotropic} 
\end{equation}
in which $\openone$ is the identity matrix;
the matrix elements in Eq.~(\ref{eq:det_matrix_angular_eq_anisotropic}) are
\begin{equation} 
\left\langle
l {\bf m}
|
M (\gamma, \lambda)
|
l' {\bf m}'
\right\rangle
 = [l(l+2 \nu ) +
\gamma] 
\,  
\delta_{ll'}  \delta_{{\bf m} {\bf m}'} 
- \lambda
\sum_{l'', {\bf m}''}
I_{ l {\bf m}, l' {\bf m}'; l'' {\bf m}'' }
\,
F_{l'' {\bf m}''} \; ,
\label{eq:matrix_elements_angular_eq_anisotropic} 
\end{equation}
where 
\begin{equation}
I_{ l {\bf m}, l' {\bf m}'; l'' {\bf m}'' }
=
\int
d \Omega_{d-1}
Y^{*}_{l {\bf m}}( {\bf \Omega}) 
Y_{l'' {\bf m}''}( {\bf \Omega}) 
Y_{l' {\bf m}'}( {\bf \Omega}) 
\;  .
\end{equation}
Finally, the components 
$\Xi_{l {\bf m}} $ of the angular wave function
can be formally obtained 
for every eigenvalue $\gamma$
in the usual way,
and satisfy
[from Eq.~(\ref{eq:wf-ultraspherical_anisotropic_normalization})]
\begin{equation}
\sum_{l, {\bf m}} 
| \Xi_{l {\bf m}} |^{2}
=
1
\; .
\label{eq:wf-ultraspherical_anisotropic_normalization_components}
\end{equation} 

As an example of this general theory,
one can consider the particular three-dimensional
case ($\nu =1/2$)
of the electron-polar molecule interaction described in Sec.~\ref{sec:dipole_bound_anions}.
In this case,
the matrix 
elements~(\ref{eq:matrix_elements_angular_eq_anisotropic})
become
\begin{eqnarray} 
\left\langle
l { m}
|
M (\gamma, \lambda)
|
l' { m}'
\right\rangle
& = &
[l(l+1) +
\gamma] 
\,  \delta_{ll'} \, \delta_{mm'} 
- \lambda 
\left\{ 
\sqrt{
\frac{(l+m)(l-m)}{(2l-1)(2l+1)}
} 
\, \delta_{l',l-1} \, \delta_{mm'} 
\right.
\nonumber
\\
& + & 
\left.
\sqrt{ 
\frac{(l+m+1)(l-m+1)}{ (2l+1)(2l+3)}} 
\, \delta_{l', l+1} \, \delta_{mm'} 
\right\} 
\; ,
\end{eqnarray}
which correspond
to a matrix of block-diagonal form
with respect to $m$ and tridiagonal in $l$.
Then, the secular determinant~(\ref{eq:det_matrix_angular_eq_anisotropic})
factors out in the form
$D(\gamma, \lambda) = \Pi_{m} D_{m}(\gamma, \lambda) $, 
with the reduced determinant $D_{m}(\gamma, \lambda) $ 
in the $m$ sector; thus,
for given $m$, the equation $\det M (\gamma,\lambda) =0$ implies that
\begin{equation}
D_{m}(\gamma, \lambda) 
=
\left|
\begin{array}{cccc}
\gamma 
&  -\frac{\lambda}{\sqrt{3}} \, \sqrt{1-m^{2}} 
& 0 
& \cdots  
\\
-\frac{\lambda}{\sqrt{3}} \, \sqrt{1-m^{2}} 
& (2 + \gamma) 
& - \frac{ \lambda}{\sqrt{15}} \,  \sqrt{4-m^{2}} 
& \cdots 
\\
0 
& - \frac{\lambda}{\sqrt{15}} \,  \sqrt{4-m^{2}} 
& (6 + \gamma) 
& \cdots 
\\
\cdots & \cdots & \cdots & \cdots
\end{array}
\right| 
=
 0 \; .
 \label{eq:electron_polar-molecule_determinant}
\end{equation}
Equation~(\ref{eq:electron_polar-molecule_determinant})
has been used for the determination of the
critical dipole moment
 $\lambda^{(*)} 
\approx 1.279$~\cite{molecular_dipole_anomaly}
when $\gamma = \gamma^{(*)}$.
When the determinant is expanded (to high orders), 
additional roots appear for the critical condition
$\gamma^{(*)} =1/4$
and for different values of $m$.
This pattern also illustrates how one would completely solve the 
generic anisotropic problem:
Eq.~(\ref{eq:electron_polar-molecule_determinant})
or its generalization (\ref{eq:det_matrix_angular_eq_anisotropic})
can be used to obtain the eigenvalues 
 $\gamma $ 
that correspond 
to a given coupling $\lambda$;
these eigenvalues replace the usual angular momentum numbers.
In the molecular physics case described above,
the  values of $\gamma $ 
 can be easily evaluated
numerically.
When $\lambda <  \lambda^{(*)} $,
no such values produce binding;
a first ``binding eigenvalue'' $\gamma_{0,0}$ is
obtained when
$\lambda \geq 
\lambda^{(*)} =
1/4$, for the first root with $m=0$;
as the strength $\lambda$
of the interaction increases,
a second  binding
eigenvalue $\gamma_{0,1}$ is produced for the first root with
   $m=1$, when
$\lambda \approx $ 7.58 or $p \approx 9.63$ D; 
the next  eigenvalue
$\gamma_{1,0}$  
arises from the second root with $m=0$; etc. 
For {\em each\/} one of these values of $\gamma = \gamma_{j, m} $,
an energy spectrum of conformal states is governed by
Eq.~(\ref{eq:cutoff_BS_regularized_energies_phenomenological}),
with 
$\Theta_{\rm eff} $ given by 
Eq.~(\ref{eq:Theta_effective_dipole}).
These bound states have been observed experimentally~\cite{mea:84,des:94}
for the case when $\gamma_{0,0}$
is the only binding eigenvalue,
a condition that corresponds to typical molecular dipole moments.

Most importantly, this analysis confirms that the conformal anisotropic problem
can be reduced to the isotropic one,
and the same symmetry-breaking considerations apply.

\section{Dimensionalities and Interdimensional Dependence}
\label{sec:dimensionalities_and_interdimensional}

 The {\em spatial dimensionality\/}
$d_{\rm eff}$ of a physical realization
of conformal quantum mechanics
is best characterized or defined as the dimension of the
configuration space needed for a complete description of the dynamics 
within the conformal approximation.
Typically, this quantity can be directly identified from the nature
of the radial variable used in the description of scale and conformal 
symmetries.

For instance,
with this convention, 
molecular anions can be naturally
seen as a three-dimensional realization
($d_{\rm eff}=3$);
the Efimov effect, in a $d$-dimensional one-particle space,
as a $(2d)$-dimensional realization 
($d_{\rm eff}=2d$); and the near-horizon conformal
physics of black holes, in $D=d+1$ spacetime dimensions, as a 
$d$-dimensional realization ($d_{\rm eff}=d$).

Of course, there is a certain degree of arbitrariness in the selection of 
$d_{\rm eff}$, due to the existence of a formal
relationship connecting problems 
of different dimensionalities.
This can be seen from the reduced Schr\"{o}dinger-like 
radial equation of a conformal 
problem~(\ref{eq:ISP_Hamiltonian_unregularized}),
\begin{equation}
\frac{d^{2}  u(r)}{dr^{2}} +
\left[
 k^{2}
+ \frac{\lambda - (l+ \nu)^{2} + 1/4}{r^{2}}
\right]
 u(r) = 0
\; .
\label{eq:radial_conformal_isotropic_eq}
\end{equation}
 Equation~(\ref{eq:radial_conformal_isotropic_eq}) depends on the number of
spatial dimensions only through the combination $l+ \nu$,
a property 
known as interdimensional dependence~\cite{van:73}.
As a consequence,
the radial part of the
solutions for any two conformal problems
 are identical when their coupling constants are
related by
\begin{equation}
\lambda (d';l') 
=
\lambda (d;l)
+
(l' - 1 + d'/2 )^{2}
-
(l - 1 + d/2 )^{2}
\;  . 
\label{eq:interdimensional_equivalence}
\end{equation}
Moreover,
\begin{equation}
\Theta (d')
 =
\Theta (d)
\;  
\end{equation}
is a {\em dimensional invariant\/}
of these formal transformations. Correspondingly,
the conformal physics is totally determined by the 
invariant value of this parameter.

However, the interdimensional equivalence of
Eq.~(\ref{eq:interdimensional_equivalence})
is severely limited 
by the fact that the full-fledged solutions (wave functions)
are not identical, because the angular momenta 
are different in different dimensionalities.
The only exception to this
is the formal equivalence among the $l=0$
angular momentum channels of problems with arbitrary  dimensionalities
(as these channels do not involve additional dimension-dependent angular 
variables); in particular,
an effective one-dimensional coupling can always be introduced 
for a $d$-dimensional problem with $l=0$:
\begin{equation}
\lambda (d'=1;l'=0) 
=
\lambda (d;l=0)
+
\frac{1}{4}
- \frac{ (d-2)^{2} }{4}
\;  . 
\label{eq:reduction_to_1D}
\end{equation}

Even in the special 
case of the equivalence described by Eq.~(\ref{eq:reduction_to_1D}),
the full-fledged wave functions still retain a trace
of the ``physical dimensionality'' $d$,
because (with an obvious notation)
$u(r) \equiv \Psi|_{d=1} (r) = 
r^{(d-1)/2}
 \Psi|_{d} (r)$;
for example,  
in the case of the three-dimensional Efimov effect,
the full-fledged wave functions are of the form $\Psi (r)
\propto r^{-5/2} u(r)$, reflecting the fact that
$d_{\rm eff}= 6$.

The example of the near-horizon conformal behavior of black holes 
presents a number of peculiar features that deserve a separate treatment
in Appendix~\ref{sec:NH_conformal_behavior}.

\section{Near-Horizon Conformal Behavior of Black Holes}
\label{sec:NH_conformal_behavior}

In this appendix we present an algebraic derivation
of the conformal invariance exhibited near the horizon
of a black hole.

From Eqs.~(\ref{eq:RN_metric}) and
(\ref{eq:massless_scalar_action}),
it follows that the equation of motion satisfied by the scalar field in the
black-hole gravitational background is
\begin{eqnarray}
\left( \Box - m^{2}  \right) \Phi
& \equiv &
\frac{1}{ \sqrt{-g} }
\partial_{\mu} \left(
\sqrt{-g} 
\,
g^{\mu \nu}
\,
\partial_{\nu} \Phi
\right)
- m^{2} \Phi
\nonumber
\\
& &
=
- \frac{1}{f} \ddot{ \Phi}
+ f \Phi''
+ \left(f' + \frac{(D-2) f}{r} \right)
\Phi'
+ \frac{1}{r^{2}} 
\triangle_{D-2} 
\Phi - m^{2} \Phi =0
\; ,
\label{eq:scalar_field_in_BH_background_1}
\end{eqnarray}
where the dots stand for time derivatives and the primes for radial derivatives
in the chosen coordinate description of the background,
while $\triangle_{D-2} $ is the Laplacian on the 
unit $(D-2)$-sphere.
In addition, 
by separation of the time and angular variables,
 \begin{equation}
\Phi  \big( t,r,  {\bf \Omega} \big)
= e^{-i \omega t} 
\varphi_{l{\bf m}} (r) Y_{l{\bf m}} \big( {\bf \Omega} \big)
\; ,
\label{eq:scalar_field_separation_variables}
\end{equation}
Eq.~(\ref{eq:scalar_field_in_BH_background_1}) turns into
\begin{equation}
 \varphi'' (r)
+ \left(
\frac{f'}{f} + \frac{(D-2) }{r} 
\right)
\varphi' (r)
+
 \left(
\frac{\omega^{2}}{f^{2} } - \frac{m^{2}}{f} - \frac{\alpha}{r^{2} \, f}
\right)
\varphi (r)
= 0
\; ,
\label{eq:scalar_field_in_BH_background_2}
\end{equation}
with $\alpha = l(l+D-3)$
being the eigenvalue of the operator
$-\triangle_{D-2} $.
Equation~(\ref{eq:scalar_field_in_BH_background_2}) 
can be further reduced, 
by means of a Liouville transformation~\cite{forsyth:Liouville}
\begin{equation}
\varphi (r) = g(r) \,
u(r) 
,
\; \; \;  \; 
g(r) = \exp \left\{ -\frac{1}{2} \int \left[ \frac{f'}{f} + \frac{(D-2)}{r} \right] dr \right\}
=
f^{-1/2} r^{-(D-2)/2}
\; ,
\label{eq:Liouville}
\end{equation}
to its normal or canonical form
\begin{equation}
u''(r) 
+
I(r) \,
u(r)
=0
\;  ,
\label{eq:scalar_field_in_BH_background_3}
\end{equation}
with normal invariant 
\begin{equation}
I(r)
=
\frac{ \omega^{2} }{f^{2}}
-
\frac{ m^{2} }{f}
-
\left[ 
\frac{ (D-2)(D-4)}{4} 
+
\frac{\alpha}{f} 
\right]
\frac{1}{r^{2}} 
- 
\frac{1}{2}
\frac{f''}{f}
+
\frac{1}{4} \frac{f'^{2}}{f^{2}}
-
\frac{(D-2) f'}{2 r f} 
\;  .
\label{eq:scalar_field_in_BH_background_3_invariant}
\end{equation}

The conformal behavior of the
Schr\"{o}dinger-like equation~(\ref{eq:scalar_field_in_BH_background_3})
near the horizon can be studied by
means of an
expansion in the variable 
\begin{equation}
x= r -r_{+}
\; ,
\end{equation}
with $r=r_{+}$ being the largest root of $f(r)=0$.
The {\em nonextremal\/} case is characterized by the condition
\begin{equation}
f'_{+} 
\equiv f'(r_{+}) 
\neq 0
\; ,
\label{eq:extremality}
\end{equation}
equivalent to
$r_{+} \neq r_{-}$.
Then,
\begin{equation}
f(r) = 
f'_{+}  \, x 
\left[ 1 + O(x) \right]
,
\; \; \;
f'(r) = 
f'_{+} 
\left[ 1 + O(x) \right]
,
\; \; \;
f''(r) = 
f''_{+} 
\left[ 1 + O(x) \right]
\; ,
\end{equation}
where 
$f''_{+} 
\equiv f''(r_{+}) $.
Thus, with corrective multiplicative factors  of the order $[1+ O(x)]$, it follows that
$f''/f  \sim f''_{+}/(f'_{+}x) $ and 
$f'/f \sim 1/x$, while $r \sim r_{+}$,
so that
the only {\em leading\/} 
terms in Eq.~(\ref{eq:scalar_field_in_BH_background_3_invariant}) are 
$\omega^{2}/f^{2} 
\sim \omega^{2}/(f'_{+} x )^{2} $
and $f'^{2}/(4f^{2} ) \sim 1/(4 x^{2})$. As a result,
Eq.~(\ref{eq:scalar_field_in_BH_background_3})  
is asymptotically 
reduced to the conformally
invariant form
\begin{equation}
u''(x)
+
\left[ 
\frac{1}{4}
+
\frac{ \omega^{2} }{(f'_{+})^{2}}
 \right]
x^{-2}
\, 
\left[ 1 + O(x) \right]
u (x)
=
0
\;  ,
\label{eq:scalar_field_in_BH_background_conformal}
\end{equation}  
where, by abuse of notation,
 we have replaced 
$u(r)$ by $u(x)$.
Equation~(\ref{eq:scalar_field_in_BH_background_conformal})
 indicates 
the existence of an {\em asymptotic conformal symmetry\/}
driven by the effective interaction
\begin{equation}
V_{\rm eff} (x) = 
- 
\frac{ \lambda_{\rm eff} }{x^{2}}
\, ,  \; \; \; \; 
\lambda_{\rm eff} = \nu^{2} + \Theta_{\rm eff}^{2}
\, , \; \; \; \; 
\Theta_{\rm eff}^{2}
= 
\bigg[
\frac{\omega}{  f'(r_{+}) } 
\bigg]^{2}
\; ,
\label{eq:conformal_interaction}
\end{equation}  
as follows by rewriting Eq.~(\ref{eq:scalar_field_in_BH_background_conformal})
in the $d$-dimensional format of Eq.~(\ref{eq:radial_conformal_isotropic_eq}).
This proves the claims made in Sec.~\ref{sec:NH_black_holes}
 and, in particular,
 Eqs.~(\ref{eq:ISP_BH_near_horizon})
and (\ref{eq:coupling_ISP_BH_near_horizon}).

A final remark is in order.
The effective Hamiltonian~(\ref{eq:scalar_field_in_BH_background_conformal})
did not fall ``automatically'' within 
the $d$-dimensional format of Eq.~(\ref{eq:radial_conformal_isotropic_eq}).
The extra terms $-[(l+ \nu )^{2} - 1/4)]/r^{2}$, usually obtained 
by reduction
of a multidimensional  Schr\"{o}dinger equation 
in flat space, are still present, but at higher orders in the expansion
with respect to the near-horizon coordinate $x$;
in Eq.~(\ref{eq:scalar_field_in_BH_background_3_invariant}),
they correspond to
\begin{equation}
-
\left[ 
\frac{ (D-2)(D-4)}{4} 
+
\frac{\alpha}{f} 
\right]
\frac{1}{r^{2}} 
=
-
\left[ 
\frac{
(l+ \nu )^{2} 
}{f} 
- \frac{1}{4}
+ \nu^{2}
\left( 1 - \frac{1}{f} \right)
\right]
\frac{1}{r^{2}} 
= O\left(\frac{1}{x}\right)
\end{equation}
[with $\nu = (d-2)/2= (D-3)/2$].
Thus, the angular momentum--together with its associated dimensionality 
variable---decouples from
the conformal interaction~(\ref{eq:conformal_interaction}) in the near-horizon
limit.
It should be noticed
that we had to rewrite Eq.~(\ref{eq:scalar_field_in_BH_background_conformal})
in the $l=0$, $d$-dimensional format in order to present this problem 
within our unified conformal model~(\ref{eq:ISP_Hamiltonian_unregularized}).
Alternatively, one could write
Eq.~(\ref{eq:conformal_interaction}) in a simpler
one-dimensional reduced form
[from Eq.~(\ref{eq:reduction_to_1D})],
$\lambda (d=1)
= \lambda_{\rm eff} - \nu^{2} + 1/4
= 
  \Theta_{\rm eff}^{2} + 1/4
$, with the same value for the
dimensional invariant  $ \Theta_{\rm eff}$.

\section{Generalized Lommel Integrals}
\label{sec:generalized_Lommel_integrals}

 In this appendix we derive a generalization of the 
 Lommel integrals~\cite{wat:44}
for an arbitrary Sturm-Liouville problem
\begin{eqnarray}
\hat{L}_{x} v(x)
=
\mu 
\varrho (x) 
v(x)
\; ,
\label{eq:Sturm_Liouville_eq}
\\
\hat{L}_{x} 
=
- \left\{
\frac{d}{dx} 
\left[ p(x) 
\frac{d}{dx} 
\right]
+ 
q(x)
\right\}
\; ,
\label{eq:Sturm_Liouville_operator}
\end{eqnarray}
and apply it to 
 the reduced 
radial Schr\"odinger equation~(\ref{eq:reduced_Schrodinger_v}).
These generalized integrals
are needed for the exact  evaluation of expectation values
in the anomaly calculation.

In what follows, 
we rewrite
the differential equation~(\ref{eq:Sturm_Liouville_eq})
in the form
\begin{equation}
\frac{d}{dx} 
\left[ p(x) 
v'(x) 
\right]
=
-
\left[
\alpha^{2}
\varrho (x) 
+ q(x)
\right]
\,
v (x)
\; ,
\label{eq:Sturm_Liouville_eq_explicit}
\end{equation}
with an eigenvalue 
$\mu  = \alpha^{2}$
and where the prime stands for  a
derivative with respect to $x$;
moreover,
 $v(x)$ can be chosen to be a real function.
Next, after
conveniently multiplying 
both sides 
by $2 p(x) v'(x)$,
and integrating them
with respect to $x$,
Eq.~(\ref{eq:Sturm_Liouville_eq_explicit})
turns into
\begin{equation}
\left. 
\left[ 
p(x) 
v'(x) 
\right]^{2}
\right|_{x_{1}}^{x_{2}}
=
-
\int_{x_{1}}^{x_{2}}
 dx
\,
p(x)
\left[
\alpha^{2}
\varrho (x) 
+ q(x)
\right]
\,
\frac{d}{dx} 
\left[ v (x) \right]^{2}
\; ,
\label{eq:Sturm_Liouville_eq_transitional}
\end{equation}
in which both the lower ($x_{1}$) and upper limits ($x_{2}$) are completely arbitrary.
Finally, after integration 
by parts and rearrangement of terms, 
Eq.~(\ref{eq:Sturm_Liouville_eq_transitional}) 
leads to 
\begin{eqnarray}
\alpha^{2} 
\int_{x_{1}}^{x_{2}}
 dx
\,
\left[ p(x) \varrho (x) \right]'
\left[
v(x)
\right]^{2}
& = &
\left.
\left[
v(x)
\right]^{2}
\,
\left\{
\left[ p(x) \frac{v'(x)}{ v(x) } \right]^{2}
+
p(x) 
\,
\left[
\alpha^{2} \varrho (x) + q(x) \right]
\right\}
\right|_{x_{1}}^{x_{2}}
\nonumber
\\
& - &
\int_{x_{1}}^{x_{2}}
 dx
\,
\left[ p(x) q (x) \right]'
\left[
v(x)
\right]^{2}
\; ,
\label{eq:Sturm_Liouville_eq_integral}
\end{eqnarray}
which generalizes the well-known second 
Lommel integral~\cite{wat:44}
of the theory of Bessel functions.
A similar procedure could be applied for a generalization of the first
Lommel integral, but this is not needed for the present purposes.

The  integral relation~(\ref{eq:Sturm_Liouville_eq_integral})
can be rewritten in a convenient form for
the reduced 
radial Schr\"odinger equation~(\ref{eq:reduced_Schrodinger_v}),
which is of the generalized Bessel form
\begin{equation}
\left\{
\frac{d^{2}}{dx^{2}}
+
\frac{1}{x} 
\,
\frac{d}{dx}
+
\left[ \alpha^{2} - {\mathcal W} (x) \right]
\right\}  v (x)
= 0
\; .
\label{eq:generalized_Bessel_eq}
\end{equation}
This is a particular 
case of the Sturm-Liouville equation~(\ref{eq:Sturm_Liouville_eq_explicit}),
with
density function
$\varrho (x) = x$, 
$p(x)=x$, and $q(x) = - x \, {\mathcal W} (x)$; however, it is also true that
 a straightforward
set of two Liouville transformations~\cite{forsyth:Liouville}
makes
Eqs.~(\ref{eq:Sturm_Liouville_eq}) 
and (\ref{eq:generalized_Bessel_eq})
formally equivalent to each other.
For  Eq.~(\ref{eq:generalized_Bessel_eq}),
$\left[ p(x) q (x) \right]' = - \left[ x^{2}  {\mathcal W} (x) \right]'$,
and
the final term in Eq.~(\ref{eq:Sturm_Liouville_eq_integral})
can be 
evaluated 
with the help of
\begin{equation}
\frac{d}{dx}
\left[
{\mathcal W} (x) \,
x^{2}
\right]
=
2 x \,
\left(
\openone
+
\frac{1}{2}
\,
{\mathcal E}_{x}
 \right)
{\mathcal W} (x)
\; ,
\end{equation}
where 
  $\openone$
is the identity operator and ${\mathcal E}_{x} = x \partial/\partial x$,
as in Sec.~\ref{sec:regularization_renormalization}.
As a consequence,
Eq.~(\ref{eq:Sturm_Liouville_eq_integral})
becomes
\begin{eqnarray}
\alpha^{2}
\,
\int_{x_{1}}^{x_{2}}
dx \,
x
\left[
v(x)
\right]^{2}
& = &
\left.
\frac{1}{2}
\,
\left[
v(x)
\right]^{2}
\,
\left\{
\left[
{\mathcal L}  (x) \right]^{2} 
 +
\left[
(\alpha x)^{2}
 - 
x^{2}
{\mathcal W} (x)
\right]
\right\}
\right|_{x_{1}}^{x_{2}}
\nonumber
\\
& + &
\int_{x_{1}}^{x_{2}}
dx \,
x
\left[
v(x)
\right]^{2}
\,
\left(
\openone
+
\frac{1}{2}
\,
{\mathcal E}_{x}
 \right)
{\mathcal W} (x)
\; ,
\label{eq:generalized_Lommel_integral}
\end{eqnarray}
where 
 ${\mathcal L}  (x)
= x v'(x)/v(x)$ and both limits are still arbitrary.
Equation~(\ref{eq:generalized_Lommel_integral}) is the desired
generalization that can be directly applied to the reduced
Schr\"odinger equations~(\ref{eq:reduced_Schrodinger_v})
and (\ref{eq:reduced_Schrodinger_v_interior})
to derive 
Eqs.~(\ref{eq:Lommel_for_mathcal_K_integral})
and (\ref{eq:Lommel_for_mathcal_J_integral}),
as we will show next. 

First,
for the interior problem ($r \leq a$),
Eq.~(\ref{eq:generalized_Lommel_integral}) 
turns into Eq.~(\ref{eq:Lommel_for_mathcal_J_integral}),
by means of the substitutions
\begin{equation}
x =  r
\, ,
\; \; \; \; 
\alpha =\tilde{k}
\, ,
\; \; \; \; 
x^{2} \, {\mathcal W} (x) 
=(l+\nu)^{2}+ r^{2} {\mathcal U} (r)
\, ,
\; \; \; \; 
v(x) = 
 w_{l+\nu}( \tilde{k} r; \tilde{k})
\, ,
\; \; \; \; 
z = \tilde{k} r
\;  ,
\end{equation}
and with integration interval
$z \in [0, \tilde{\xi}]$, where $\tilde{\xi} =
\tilde{k} a $.
For this case,
when 
$ r^{2} {\mathcal U} (r) \rightarrow 0$,
that is, for regular core potentials,
the behavior of the differential equation
at the origin implies that 
the contribution from the first term on the right-hand side of
Eq.~(\ref{eq:generalized_Lommel_integral})
is zero for $r=0$.

Second, in a similar manner,
for the exterior problem ($r \geq a$),
Eq.~(\ref{eq:generalized_Lommel_integral}) 
turns into 
Eq.~(\ref{eq:Lommel_for_mathcal_K_integral}),
by means of the substitutions
\begin{equation}
x = r
\, ,
\; \; \; \; 
\alpha  = k = i \kappa
\, ,
\; \; \; \; 
x^{2} \, {\mathcal W} (x) 
=(l+\nu)^{2} - \lambda = - \Theta^{2}
\, ,
\; \; \; \; 
v(x) = K_{i \Theta} (\kappa r)
\, ,
\; \; \; \; 
z = \kappa r
\;  ,
\end{equation}
and with integration interval
 $z \in [\xi, \infty]$,
with $
\xi = \kappa a
$.  Here,
the behavior of the differential equation
at infinity implies that 
the contribution from the first term on the right-hand side of
Eq.~(\ref{eq:generalized_Lommel_integral})
is also zero at that point.


\begin{thebibliography}{99}
\bibitem{BH_thermo}
The black-hole entropy in $d$ dimensions
is proportional to the ``area'' of the outer-horizon
hypersurface:
Ref.~\cite{mye:86}.
\end{thebibliography}
\end{document}